\documentclass[aps,pra,twocolumn,showpacs,groupedaddress]{revtex4}
 
\usepackage{bm}% bold math
\usepackage{amsmath,graphicx}

\begin{document}

\title{Resonant tunneling in a schematic model}

\author{Sonia Bacca\footnote{s.bacca@gsi.de} and Hans Feldmeier\footnote{h.feldmeier@gsi.de}}

\affiliation{Gesellschaft f\"ur Schwerionenforschung, Planckstr.~1, 64291 Darmstadt, Germany}

\date{\today}

\begin{abstract}
Tunneling of an harmonically bound  two-body system through  an
external Gaussian barrier is studied in a  
schematic model which allows for a better understanding of intricate
quantum phenomena.
 The role of finite size and internal structure is investigated in a
 consistent treatment.
The excitation of 
internal degrees of freedom gives rise to a peaked structure in the
penetration factor. The model results indicate that for soft systems
the adiabatic limit is not necessarily reached although often assumed in
fusion of nuclei and in  electron
screening effects at astrophysical energies. 
\end{abstract}

\pacs{03.65.-w, 03.65.Xp, 24.10.Eq, 25.70.Ef}

\maketitle

The phenomenon of  quantum tunneling is relevant in
several areas of physics,  from chemical reactions,  electronic
circuits, to nuclear fission and fusion processes below the Coulomb
barrier.
In many cases one has to deal with the tunneling of  composite
objects. 
In general one tries to identify  macroscopic degrees of freedom
(denoted by $R$)  and intrinsic ones (denoted by $\xi$), decomposing the Hamiltonian accordingly:
\begin{equation}
H= H_{mac}(R)+ H_{int}(\xi)+ V(R,\xi)\ .
\end{equation}
The macroscopic part, $ H_{mac}(R)$, is the Hamiltonian for the (few)
collective variables and contains besides the collective kinetic
energy the macroscopic potential that has to be transversed by
quantum tunneling. $H_{int}(\xi)$ contains the (many) intrinsic
degrees of freedom of the many-body system, while $V(R,\xi)$ describes
an interaction between intrinsic and macroscopic variables. For
example,  if one
wants to describe electron screening effects at astrophysical energies \cite{astro-ref}, $R$ could be the distance between the nuclei of two fusing
atoms  and $\xi$ could denote the electron degrees of freedom. Or in
nuclear physics $R$ may denote the shape degrees of freedom for a
heavy nucleus that undergoes  spontaneous fission and  $\xi$ the individual 
nucleon degrees of freedom (see for example \cite{fission-ref}).
The question of how the
tunneling reaction is influenced by finite size and structure of the
composite object is an intriguing  one.
An important aspect is to determine which degree of freedom must be taken into account in a theoretical
description and which can be neglected.
The description of the
tunneling of a composite object with many degrees of freedom
is a very complex many-body problem. Therefore, very often radical
approximations are made.
As an example, nuclear cross
sections at very low energy
which are important in astrophysics and where no  experimental data
exist, are often determined by extrapolations based on the
one-dimensional result for
point-like particle tunneling.
On the other hand, in heavy-ion fusion reactions it is well known that the
coupling of the relative motion of the colliding nuclei with the nuclear
intrinsic motion strongly  enhances the fusion cross section in
the sub-barrier  energy region. 
This was proven in a number of precise experiments, where the fusion
cross section for intermediate mass systems (see
e.g. \cite{fusion_exp})  was found to be higher than simple
one-dimensional prediction for tunneling through a potential barrier
formed by the attractive nuclear interaction and the repulsive Coulomb
force. The coupling between macroscopic and microscopic coordinates
 has been studied in
multidimensional approaches (with many internal degrees of freedom), with different coupling schemes and with
different approximations (see \cite{review} and references therein).

In this paper we investigate  the relation between the translational
motion and the internal degrees of freedom of a composite object in a schematic but fully consistent model.
In spite of  the simplicity of the model it has all the ingredients to
understand for example the   sub-barrier fusion of soft  nuclei, that
can easily vibrate. 
In fact the low energy fusion cross section is usually dominated by
$s$-wave fusion so that one deals only with one collective variable,
the radial coordinate $R$, and a few low lying excited states \cite{review},
like in the schematic model we describe in the following.

The Hamiltonian of two interacting particles with identical mass $m$
under the influence of  an external potential
barrier is 
\begin{eqnarray}
\label{hamx1x2}
\nonumber
H(x_1,x_2)&=&-\frac{\hbar^2}{2m}\left[\frac{d^{2}}{dx^2_1}+\frac{d^{2}}{dx_2^2}
\right]+V_{int}(x_1-x_2)
\\ &+&V(x_{1})+V(x_{2}),
\end{eqnarray}
where  $x_{1,2}$ denote the coordinate of particle $1$ and $2$,
respectively, while $V_{int}$ and $V$ are the intrinsic potential
and the external barrier, felt by both particles.
Performing a transformation to the center of mass (C.M.)  and relative
coordinate  
denoted by $R$ and $\xi$, respectively, the Hamiltonian becomes 
\begin{equation}
\label{hamrxi}
H(R, \xi)=-\frac{\hbar^2}{2M}\frac{d^2}{dR^2}+H_{int}(\xi)+
V\!\!\left(\!R+\frac{\xi}{2}\!\right)+ V\!\!\left(\!R-\frac{\xi}{2}\!\right).
\end{equation}
Here
 $H_{int}(\xi)= -\frac{\hbar^2}{2\mu}\frac{d^{2}}{d\xi^2} + V_{int}(\xi)$ is the 
 Hamiltonian of the intrinsic system, while  $M=2m$ and $\mu=m/2$ represent the center of mass
 and the reduced mass, respectively. As one can see, the external
 potential, depending on its functional form, may generate  coupling terms
 of different orders  between the macroscopic and the
 internal coordinates.
%
% Equation ~(\ref{hamrxi}) can be more generally regarded as a Hamiltonian
% describing the kinetic energy of a macroscopic coordinate $R$, the
% motion of a compound system  
%with intrinsic coordinate $\xi$ (or with a set of  intrinsic  coordinates $\xi=\{ \xi_i \}$) and a coupling term between $R$ and
%$\xi$, which in this case is  introduced by the external barrier. In
%this way one could more generally tackle the tunneling problem of a
%more complex system in a three-dimensional space.
%For instance, the electron screening in astrophysical reactions or the
%fusion of two complex nuclei could be described by such an Hamiltonian
%as well \cite{review}.

The standard theoretical approach to study the effect of internal excitations induced by
the  coupling potential is to solve the coupled-channel
equations, which in our case read
\begin{equation}\label{coupled}
-\frac{\hbar^2}{2M}\frac{d^{2}}{dR^{2}}
\phi_{ji}(R)+\sum_{n=0}^{N} 
[ V_{jn}(R) +(\epsilon_n-E)\delta_{jn} ]
 \phi_{ni}(R)=0\ ,
\end{equation}
where $E$ is the total energy of the system and $\epsilon_n$ is the
internal excitation energy.
These equations are obtained  by introducing the eigenstates of the 
internal systems, i.e. 
\begin{equation}\label{internal}
\left(-\frac{\hbar^{2}}{2\mu}\frac{d^{2}}{d\xi^{2}}+V_{int}(\xi)\right)\chi_{n}(\xi)=\epsilon_{n}\chi_{n}(\xi),
\end{equation}
and expanding the total wave function  as
\begin{equation}
\varphi(R,\xi)=\sum_{n=0}^{N}\phi_{ni}(R)\chi_{n}(\xi),
\end{equation}
 $N$ being the number of internal excitations considered. 
 The expansion coefficients $\phi_{ni}(R)$  depend on the
C.M. coordinate. Here  the sub-index $i$ denotes the initial channel,
i.e. the starting energy level of the internal system.
In Eq.~(\ref{coupled}) the  potential matrix elements 
\begin{equation}\label{vnj}
 V_{jn}(R)=\int_{{-\infty}}^{{\infty}}\!\!\!\!d\xi~\chi_{j}^{*}(\xi) \!\left[\!V\!\left(\!R+\frac{\xi}{2}\!\right)\!+\!V\!\left(\!R-\frac{\xi}{2}\!\right)\!\right]\!\chi_{n}(\xi)
\end{equation}
can be interpreted as the  effective potentials felt by the
C.M. due to presence of internal degrees of freedom.

Equation (\ref{coupled}) consists of a set of $N$  coupled second
order differential
equations and, in case the particle is incident on the barrier from the
left hand side, the boundary conditions we require for its solution  are
\begin{eqnarray}\label{boundarycond}
\lim_{R\to-\infty}\phi_{ni}(R)&=&\delta_{ni}e^{ik_{n}R}+A_{ni}e^{-ik_{n}R},\nonumber\\
\lim_{R\to\infty}\phi_{ni}(R)&=&B_{ni}e^{ik_{n}R}.\label{boundary}
\end{eqnarray}
Here $k_n=\sqrt{2M(E-\epsilon_n)/\hbar^2}$ is the wave number of the
$n$-th channel.
The inclusive penetration factor, or total transmission coefficient, is
then given by
\begin{equation}\label{totalT}
T=\sum_{n=0}^{N} T_{ni}\ ,
\end{equation}
where the transmission probability for each channel is defined as
\begin{equation}
T_{ni}=\frac{k_{n}}{k_{i}}\left|B_{ni}\right|^{2}.
\end{equation}

In the schematic model  that we would like to solve we assume an
harmonic internal potential
$V_{int}(\xi)=\frac{1}{2}\mu\Omega^{2}\xi^{2}-\frac{1}{2}\hbar \Omega$ and
a  Gaussian external barrier 
$
V(x)=V_{0}e^{-\nu^{2}x^{2}}.$
In this  case the internal excitation energy becomes $\epsilon_n=n\hbar
\Omega$ and the effective potential matrix elements have the following
analytical form \cite{razavy, tables}
\begin{eqnarray}
\label{vjn_razavy}
V_{jn}(R)=&{4V_{0}}\frac{\beta}{(2^{j+n}n!j!)^{1/2}}\frac{e^{-\frac{4\beta^{2}\nu^{2}R^{2}}{4\beta^{2}+\nu^{2}}}}{(4\beta^{2}+\nu^{2})^{1/2}}\\
\times&\sum\limits_{k=0}^{\mathrm{min}(n,j)}2^k k!\binom{j}{k}\binom{n}{k}\left(\frac{\nu^{2}}{4\beta^{2}+\nu^{2}}\right)^{\frac{j+n}{2}-k}\nonumber\\
\nonumber
\times&H_{j+n-2k}\left(-\frac{2\nu\beta R}{(4\beta^{2}+\nu^{2})^{1/2}}\right),
\end{eqnarray}
if $n+j$ is even, and $V_{jn}(R)=0$ otherwise, where  $\beta=(\mu \Omega/
 \hbar)^{1/2}$ is the harmonic oscillator parameter.
 We would like emphasize that in this case the potential matrix
elements can be calculated exactly and treated consistently within the 
model:  they are given by the product of a Gaussian
times a linear combination of Hermite polynomials. Therefore,  they may show
some structure depending on the chosen parameters.
This is not equivalent to making a separable ansatz for the coupling
potential and then assuming constant form factors or a Gaussian
parameterization  as proposed in Ref.~\cite{dasso-landowne}.
One can note that the potential matrix elements of (\ref{vjn_razavy})  connecting channel
$j$  with channel $n$ depend on the  internal system via
the $\beta$ parameter. 
The first matrix element, $V_{00}$, which accounts for the elastic
transition from the ground state ($j=0$) to itself ($n=0$), is  a simple
Gaussian function whose width depends on the original width of the external
barrier ($\sim 1/\nu$) and on the harmonic oscillator parameter $\beta$.  In case of a very
stiff internal harmonic oscillator, i.e. for $\Omega \rightarrow
\infty$, one sees that $V_{00}(R)\rightarrow 2V(R)$. This is
the limit case in which the composite object behaves like  a point
particle, feeling two times the external potential. Thus, for finite
$\Omega$  the diagonal matrix elements $V_{nn}$  account for
finite-size effect, while the coupling terms ($n\ne j$) are 
responsible for transition to the excited states. 
Performing a Taylor expansion of the coupling potential for $\xi \ll
1/\nu$  up to the second order one gets
\begin{equation}
\label{taylor}
V\!\left(\!R+\frac{\xi}{2}\!\right)\!+\!V\!\left(\!R-\frac{\xi}{2}\!\right)
= 2V(R)+V''(R) \frac{\xi^2}{4}+ \dots \ .
\end{equation}
In case the two-body system is composed of identical particles, both
interacting with a symmetrical external barrier, the  coupling potential does not
include any linear coupling term as $V'(R)\xi$. Actually, it consists
in an expansion over all even coupling terms.    
This means that the barrier effect on the intrinsic system is
reduced, up to the second order of the expansion, 
to an additional harmonic potential that shrinks or stretches  it, depending whether the second derivative of the
external potential  is positive or negative, respectively.
By comparing the intrinsic harmonic potential $V_{int}$ with the
induced harmonic potential $V''(R) \frac{\xi^2}{4}$, for example  at
$R=0$ where $V''(R)$ presents a minimum with maximal amplitude,
one can say 
 that if $\hbar\Omega > \hbar\nu \sqrt{V_0/\mu}$ (or  $\hbar\Omega < \hbar\nu \sqrt{V_0/\mu}$) the
intrinsic system is stiff (or soft) with respect to induced excitations. If the object is very stiff one expects the internal excitation to play a negligible role, whereas in case of a very soft system many internal levels have to be taken into account. 
\begin{figure}
\resizebox*{8cm}{9cm}{\includegraphics[angle=0]{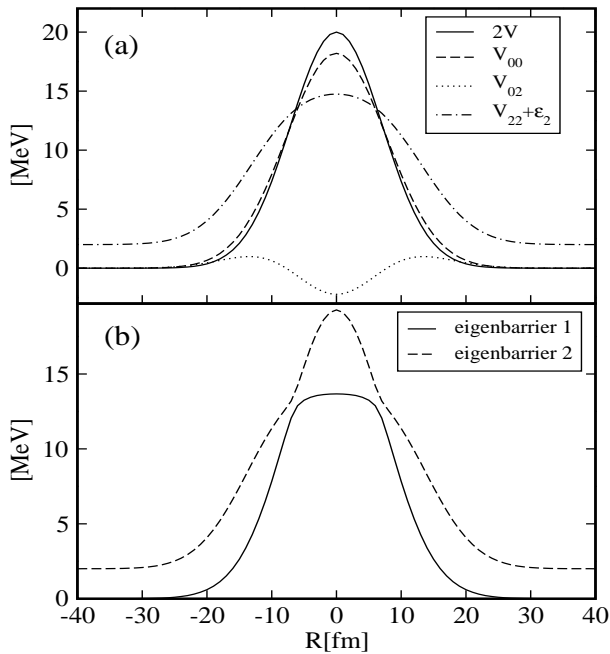}}
\caption{Effective potential matrix elements for a two level
  system in comparison with the external barrier (parameters are:
  $\hbar\Omega=$ 1 MeV, $m=$ 938 MeV, $V_0=$ 10 MeV and $\nu=$ 0.1 fm$^{-1}$).}
\end{figure}

In the following we  solve the coupled channel problem using
the potential matrix elements of Eq.~(\ref{vjn_razavy}). In order to
gain insight into the effect of the coupling  we  restrict
ourselves to the case of a two level system.  The internal degrees of
freedom are taken to be initially in the ground state with excitation energy
$\epsilon_0$.  Thus, in the presented model
the only possible transition will be to the second excited
state ($n=2$), since the coupling term $V_{01}(R)$ vanishes due to the
mirror symmetry of the Gaussian barrier. 
For our two level system we can then define the potential matrix as
following
\begin{equation}
\label{matW}
W(R)=\left(
\begin{array}{cc}
{ V_{00}(R)}&{V_{02}(R)}\\
{V_{20}(R)}&{V_{22}(R) +\epsilon_2}
\end{array}
\right),
\end{equation}
where $\epsilon_2=2\hbar \Omega$ is the
energy difference between the two levels, being $\epsilon_0=0$.
Since we are interested in studying the effect of induced internal
excitations, we will start to consider a soft object, where the
internal structure plays a relevant role.
 In our first analysis we use  following parameters: harmonic oscillator frequency
$\hbar\Omega=$ 1 MeV,  mass
$m=$ 938 MeV, barrier amplitude $V_0=$ 10 MeV, barrier inverse width
$\nu=$ 0.1 fm$^{-1}$, which lead to  $\hbar\nu \sqrt{V_0/\mu}=2.88$ MeV.

In Fig.~1 (a) we firstly show the consistent  potential
matrix elements of Eq.~(\ref{vjn_razavy}) for the chosen parameters, in
comparison with the external barrier.
One can note that, due to finite size effect, the potentials $V_{00}$
and $V_{22}$ show a broadened structure  with respect to the external
potential, while the coupling term $V_{02}$ changes sign two times.
In Fig.~1 (b), we present the  the so-called
eigenbarriers (or eigenpotentials), obtained 
  diagonalizing the potential matrix of Eq.~(\ref{matW}) at each position of the macroscopic coordinate R (see
e.g. \cite{paper2}).

We have then performed the coupled channel calculation integrating Eq.~(\ref{coupled})
from $R=-40$ fm  to $R=-40$ fm,  and imposing the incoming wave  boundary
condition of Eq.~(\ref{boundarycond}). It is known that with this method,  often used  in 
heavy-ion collisions calculations \cite{hion}, it is
 sometimes 
difficult to obtain a stable solution with a controlled numerical
accuracy.
In order to check the numerical results we also solve the problem with
the more stable method of the
variable reflection amplitude \cite{razavy}, where the set of two
coupled second order linear differential equations is transformed into a
set of four coupled non linear first order differential equations.
Other similar methods   have been proposed to stabilize the
numerical solutions of the coupled channel  problem (see
e.g. \cite{localreflmat}).  
We have obtained the same result with the above mentioned two methods with a
 relative percentage error  of about $1$-$3$ $\%$ in the  presented energy
 region, indicating
that the numerical accuracy is under control in the considered
example.

\begin{figure}
\resizebox*{8cm}{5.6cm}{\includegraphics[angle=0]{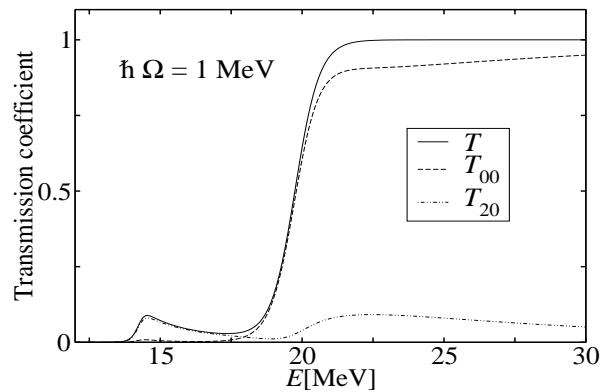}}
\caption{Total transmission coefficient and separate  $T_{20}$ and
  $T_{00}$ contribution as a function of the total energy of the
  system for $\hbar \Omega=1$ MeV.}
\end{figure}

In Fig.~2 we show the result of the total transmission coefficient $T$ and
of the separate contributions $T_{20}$ and $T_{00}$. 
As one can see, the penetration factor presents a peaked structure at
a total energy value of  about $E_{\rm peak}=14$ MeV. 
The pronounced peak is mainly due to  excitation to the 
energy level $\epsilon_2$, as can be seen from the fact that also $T_{20}$ presents a peak, while
the elastic channel $T_{00}$ is rather flat in this energy
region. This agrees with the intuitive picture that the system 
avoids the higher barrier $V_{00}$ by going to the excited state which
then 
tunnels
easily through the lower barrier $V_{22}+\epsilon_2$ (see Fig.~1).
However, this picture does not hold for higher energies where the situation is inverted due to the
effect of the coupling and finally,  for  energies somewhat higher than the
external barrier ($20$ MeV),  the elastic channel $T_{00}$ dominates
the total transmission. Thus, a proper treatment of the coupling
$V_{02}$ is important.

We would like to point out that  the consistent treatment
of the internal degrees of freedom in Eq.~(\ref{vnj}) produces 
  different widths for the diagonal potential matrix elements and  
oscillations in the coupling term. This leads to
the emergence of a resonant structure in the penetration factor.
A  peak in the transmission coefficient
is not found if one
parameterizes each potential matrix element with a simple Gaussian
function with the same width as done in \cite{dasso-landowne,paper2, paper3},
 where a   shoulder-like
structure is found.

\begin{figure}
\resizebox*{8cm}{5.5cm}{\includegraphics[angle=0]{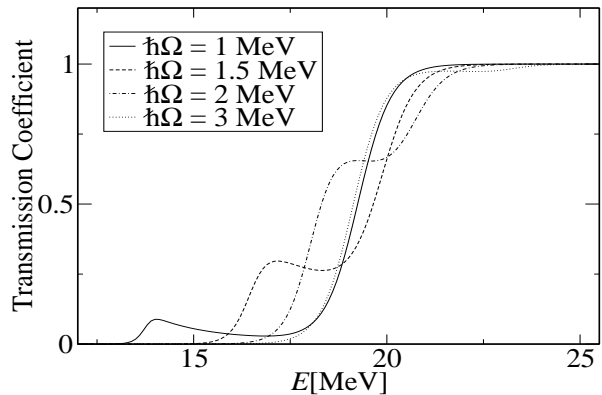}}
\caption{ Transmission coefficients for different values of the
  harmonic oscillator frequency as a function of the total energy.}
\end{figure}

We can then investigate the effect of the different stiffness of the
composite object
by keeping the same parameters for the external barrier and changing
the internal frequency of the harmonic oscillator. In Fig.~3 we 
show the  transmission coefficients for different values of $\hbar\Omega$ as a function of the total energy. One
  can see the the peaked structure moves towards higher energies with
  growing intrinsic frequency. When $\hbar\Omega$ is larger than  $\hbar\nu \sqrt{V_0/\mu}$ , the structure in $T$ becomes less
  pronounced and more similar to a shoulder, as found in \cite{paper2}.
   The reason is   that, as $\Omega$ increases, the widths of the diagonal matrix
   elements $V_{00}$ and $V_{22}$ become more similar and the
   coupling $V_{02}$ gets smaller, leading to a potential matrix
   similar to the parameterization  of Ref.~\cite{paper2} for high
   excitation energies.  
  For $\hbar\Omega=2$ MeV and higher the transmission coefficient at low
  energy is dominated by the elastic channel $T_{00}$, since the
  consistent treatment of the potential matrix elements gives a
  $V_{00}$ barrier lower than the $V_{22}+\epsilon_2$, in
  contradistinction to the case of $\hbar\Omega=1$ MeV  depicted in
  Fig.~1.

We also investigated the model with two excited intrinsic states. 
%We verified that the two level system approximation is rather good only in case of
%$\hbar\Omega=3$ MeV and higher, i.e. for a stiff object, as expected.
 In general the three level system picture within this consistent model
leads  to the emergence of two peaked structures: the position of
the first peak is shifted towards  lower energies and that of the
second towards higher energies with respect to the location of the
single  peak found in the two level system.
Such a  behavior was also found  for the shoulder-like structure of the three level system
 in \cite{paper3} with respect the two level system in
\cite{paper2}.
In case of three  levels the structure of the
consistent matrix elements  becomes more complicated, therefore we
prefer to stick to the simpler two level approach to understand further the effect of the coupling. 
  
In order to estimate the dynamical effect due to excitation of  internal
degrees of freedom one should compare the coupled channel calculation with the
uncoupled problem, where only the potential $V_{00}$ is
considered.  As already mentioned, this matrix element does not allow
for internal transitions, but still accounts for finite size effects. 
 This is different from the tunneling through the bare
 barrier $2V$, since this would be equivalent to reduce the problem to
 the tunneling of a point-like particle. 
In   Fig.~4 we show the above mentioned transmission coefficients in case of a soft object, i.e. for $\hbar\Omega=1$ MeV.
 In the peak region the effect of the excitation of the
internal degrees of freedom leads  to an enhancement of the
transmission coefficient $T$ of about 5 orders of magnitude.
On the other hand at higher energies the tunneling probability is
lower than in case of no coupling. The enhancement of  $T$ in the
coupled channel calculation with respect to the no coupling case
strongly depends on the harmonic oscillator frequency: it decreases
with growing $\Omega$, i.e. with increasing stiffness of the
composite object.
 In Fig.~4 we also compare our result with the transmission
 coefficients found for the tunneling through the eigenbarriers of the
 two-level system. 
% The tunneling through the first eigenbarrier is often
% called in literature adiabatic transmission.
 In the following we recall  the meaning of the eigenbarriers.

\begin{figure}
\begin{center}
\resizebox*{8cm}{5.5cm}{\includegraphics[angle=0]{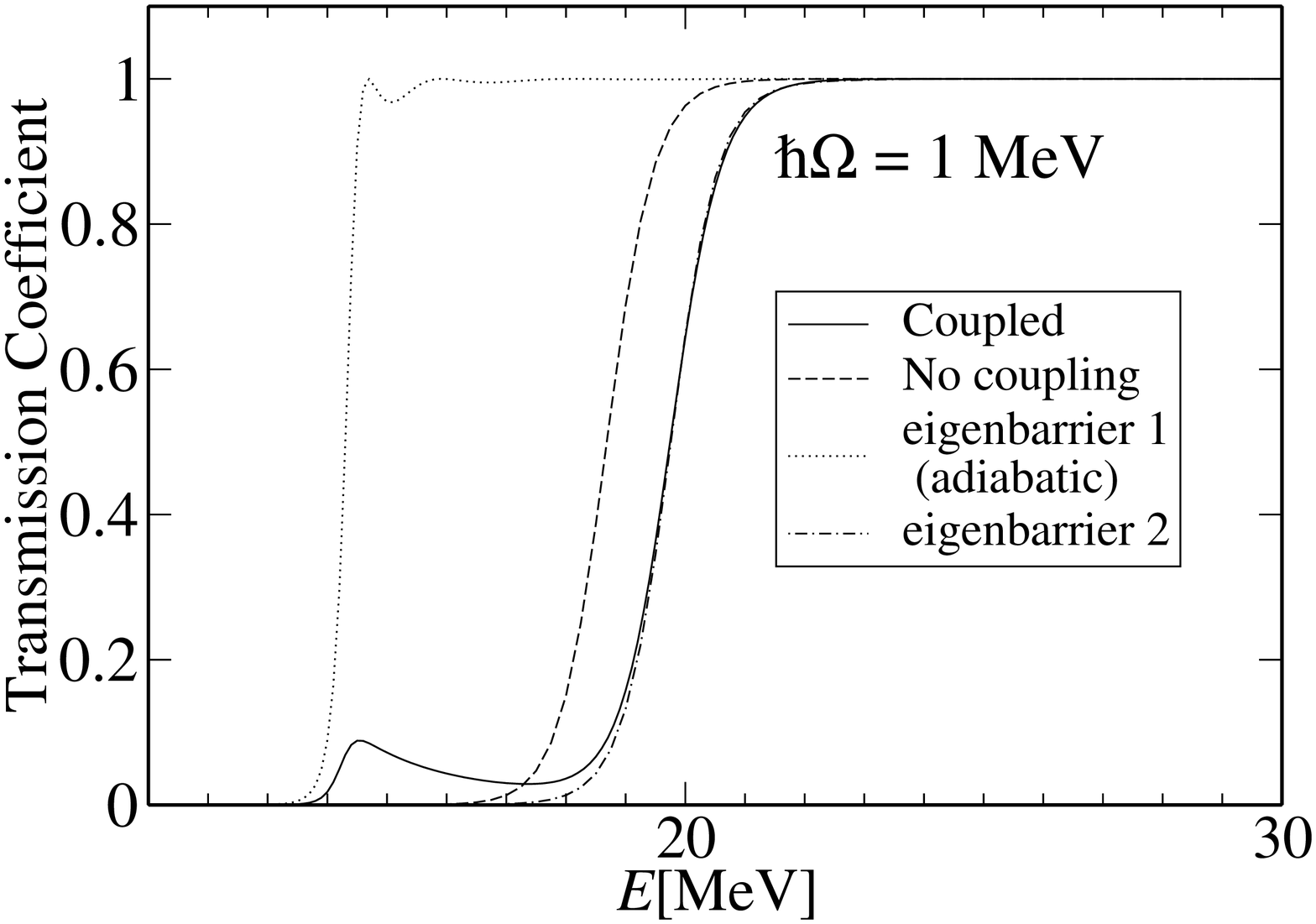}}
\caption{Effects of the internal degrees of freedom for
  $\hbar\Omega=1$ MeV: comparison
  of the coupled channel result with the no coupling
  calculation. Transmission coefficients for tunneling through the
  eigenbarriers are also shown.}
\end{center}
\end{figure}

Considering the  part of the Hamiltonian of Eq.~(\ref{hamrxi}) that
depends on the intrinsic coordinate, i.e.
\begin{equation}
\label{hamad}
\tilde{H}(R, \xi)=H_{int}(\xi)+
V\!\!\left(\!R+\frac{\xi}{2}\!\right)+ V\!\!\left(\!R-\frac{\xi}{2}\!\right)
\end{equation}
one can define the adiabatic states by minimizing it  with respect to the
internal degrees of freedom at each position $R$ \cite{review}.  
%\begin{equation}
%\label{min}
%\begin{array}{c}
%{\rm min}\\
%\vspace{-0.4cm}
%{\phi^*_{ni}}
%\end{array}
% \left\{ \!\!\left< \varphi(R,\xi) | \tilde{H}(R, \xi) | \varphi(R,\xi)
%  \right> \! - \!\lambda  \left< \varphi(R,\xi)|  \varphi(R,\xi) \right> \!\!   \right\}  =0,
%\end{equation}
%where $\phi^*_{ni}$ play the role of coefficients in the 
%   expansion of the wave function in terms of the intrinsic states of Eq.~(\ref{internal}).
This translates then into an eigenvalue problem, which in our case reads  
\begin{equation}
\label{eig}
W(R) \left(
\begin{array}{c}
\phi_0 (R)\\
\phi_2(R)
\end{array}
\right)=\lambda \left(
\begin{array}{c}
\phi_0 (R)\\
\phi_2 (R)
\end{array}
\right),
\end{equation}
with $W(R)$ as defined in Eq.~(\ref{matW}). The solution of the
eigenvalue problem is found by diagonalizing the potential matrix,
i.e. by considering the eigenbarriers.  
In literature the tunneling through the lowest
eigenbarrier is often called adiabatic transition.

As one can notice from Fig.~4, the adiabatic picture is very
different from the  result of the coupled channel calculation, though
a small structure is found in the adiabatic tunneling in the same
position as the pronounced peak of the coupled calculation. Furthermore, we
observe that the adiabatic transmission coefficient is always larger
than the coupled  result. For higher energy the coupled channel
calculation agrees with the result obtained for the tunneling through
the second eigenbarrier.

The concept of eigenbarriers is useful in case one wants to
describe the fusion cross section as given by an average over the
contribution form each eigenbarrier with appropriate weight factors.
A method to extract the barrier distribution from the measured cross
section was  proposed in Ref.~\cite{barrier_distrib}. From a
purely theoretical point of view, the barrier distribution picture is
correct only when the transformation that diagonalizes the matrix
$W(R)$ does not depend on the coordinate $R$.  An approximation which
is often made consists in considering the eigenbarriers
and then evaluating the weight factor at a fixed position of $R$,
assuming them to be constant.
 In Ref.~\cite{paper2} it was shown  that  in case one parameterizes the
potential matrix elements by  Gaussians of the same width
the weight factors are almost
constant as a function of the energy.  It was also proven that they are
very different from those approximately estimated at the location of
the maximum of the bare barrier  \cite{const_form_fctr}.
For a two level system the weight factors are given by \cite{paper2} 
\begin{eqnarray}
\nonumber
w_{-}(E)&=&[T(E)-T_{-}(E)]/[(T_{+}(E)-T_{-}(E)],\\
w_{+}(E)&=&[T_{+}(E)-T(E)]/[(T_{+}(E)-T_{-}(E)],
\end{eqnarray}
where $T_{-,+}(E)$ denote the transmission coefficient for the first
eigenbarrier and for the second, respectively, and $T(E)$ is the total
transmission as calculated in the coupled channel case.
In Fig.~5 we show these optimum weight factors in our consistent model
for the case of  $\hbar\Omega=1$ MeV.
\begin{figure}
\resizebox*{8cm}{6cm}{\includegraphics[angle=0]{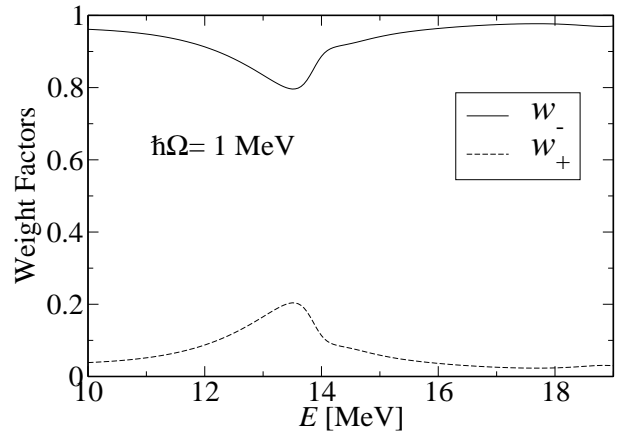}}
\caption{Weight factors for the two eigenbarriers in case of
  $\hbar\Omega=1$ MeV.}
\end{figure}
One can note that they are not constant in the energy region where the
peaked structure in $T$ is presented, showing that  the
approximation of constant weight factors does not hold.  This is related to the fact that the
transition is not adiabatic, as already discussed. We observe that
with increasing stiffness of the composite object the weight factor
show a less pronounced variation as a function of the energy. 

In order to gain more insight into the meaning of the adiabatic
picture, we recall that in case of a two level system the
orthogonal rotation that is needed to diagonalize the symmetric matrix
$W(R)$ can be parameterized by one single mixing angle $\theta(R)$ as
follows
\begin{equation}
{\mathcal R}_{\!\theta(\!R\!)}=\left(
\begin{array}{cc}
\cos \theta(\!R) & \sin \theta(\!R) \\
-\sin \theta(\!R) & \cos \theta(\!R) 
\end{array}
\right).
\end{equation} 
At this point, if one would like to transform the coupled channel
equation of (\ref{coupled}), one should
accordingly transform also the kinetic energy, which does not commute
with the rotation  operator, that depends on the coordinate $R$.
 In fact, denoting
with $K$ the diagonal kinetic energy matrix, the
transformed matrix  looks like
\begin{equation}
\label{kintras}
{\mathcal R}^T_{\!\theta(\!R\!)} K {\mathcal R}_{\!\theta(\!R\!)} \!\! = \!\!\frac{\hbar^2}{2M}\! \! \left(\!\!\!
\begin{array}{cc}
\!-\frac{d^2}{dR^2}\!+\! (\!\theta'(\!R)\!)^2 \!\! & - \theta''(\!R) \!-\!2 \theta'(\!R) \frac{d}{dR}  \\
 \theta''(\!R)\!+\!2 \theta'(\!R) \frac{d}{dR}\!\! &\! -\frac{d^2}{dR^2}\!+\! (\!\theta'(\!R)\!)^2
\end{array}
\!\!\!\right)\!.
\end{equation} 
 Thus, one can see that the adiabatic picture, which consists in
 considering only the first eigenbarrier solving an uncoupled problem,  is valid in
 case that  the first and second derivative of the
 mixing angle are negligible, i.e. in case that the transformed
 kinetic energy tends to the original diagonal one.
In order to see whether this is the case or not in the
considered example we show in Fig.~6 the
mixing angle (a) and its derivative (b) for different values of the internal
frequency $\Omega$ as a function of the C.M. coordinate.
\begin{figure}
\resizebox*{8cm}{8cm}{\includegraphics[angle=0]{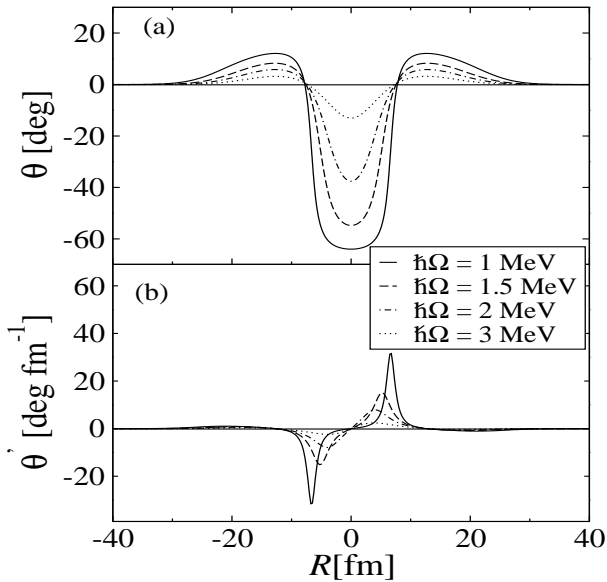}}
\caption{Mixing angle and its first derivative as a function of
  C.M. coordinate for different values of $\hbar \Omega$.}
\end{figure}
 Firstly
one notes that in all cases $\theta$ crosses  zero two times; these are the two points
in which the coupling matrix element $V_{02}(R)=0$ and thus $W(R)$ is
already diagonal. This is not the case if one parameterizes all matrix
elements with  Gaussian functions of the same width as was done in
\cite{dasso-landowne, paper2}, where the mixing angle is always
positive and can be maximal  $45^{\circ}$ in case of strong coupling.  
If one then focuses on the softer case of $\hbar\Omega=1$ MeV, which
corresponds to the example shown in Fig.~2 and Fig.~4,  one observes that the angle changes suddenly
sign, going from about $12^{\circ}$ to about  $-64^{\circ}$.
The
derivative of the mixing angle is thus very large in the vicinity
of the two zeros of $\theta$.  For this reason  its
contribution  in the transformed kinetic energy is not negligible.
 In fact, 
the extra diagonal term  $\frac{\hbar^2}{2M}(\theta'_{max})^2$, which
is about $3.25$ MeV at $R=\pm 6.6$ fm, has to be added to the
adiabatic potential, which is maximally $13.67$ MeV.
Moreover, the off-diagonal matrix elements contain the second
derivative of $\theta(R)$ which gives maximum values of
$\frac{\hbar^2}{2M}\theta''_{max}\approx 4.34$ MeV, invalidating
the adiabatic assumption.

The rapid change of the mixing angle is related to a so-called Landau-Zener  pseudo-crossing of the two levels \cite{lz}.
Denoting the rotated state with
\begin{equation}
\left(
\begin{array}{c}
\tilde{\phi}_0(R)\\\tilde{\phi}_2(R)
\end{array}
\right ) 
= {\mathcal R}_{\!\theta(\! R \!)}\left(
\begin{array}{c}
{\phi}_0\\{\phi}_2
\end{array}
\right ) 
\end{equation}
and considering the optimal case in which the mixing angle varies
suddenly as  $\theta(R_1) \rightarrow  \theta(R_2) =\theta(R_1) \pm
\pi /2 $ going from coordinate $R_1$ to $R_2$,  one can note that the
two rotated states invert each other 
in the sense
$\tilde{\phi}_0(R_1) \rightarrow \tilde{\phi}_2(R_2)$ and
$\tilde{\phi}_2(R_1) \rightarrow \tilde{\phi}_0(R_2)$. In other words,
the first energy level suddenly becomes the second and vice versa. 
In case of a soft object the variation of the mixing
angle does not give rise to the maximal pseudo-crossing, but still,
for example, at $R_1 \approx -14$ fm we have $\theta=12^{\circ}$ and
$\tilde{\phi}_0(R_1)=(0.98 \phi_0+ 0.21 \phi_2)$ and  at $R_2=0$ we
have $\theta=-64^{\circ}$ with
$\tilde{\phi}_2(R_2)=(0.90 \phi_0+ 0.44 \phi_2)$, so
that $\tilde{\phi}_2(R_1)$ and $\tilde{\phi}_2(R_2)$ are very similar.
 Therefore,  there is partial  pseudo-crossing which is related to a non adiabatic transition, as discussed above.
In fact, looking at Fig.~6 one can see that if the composite object becomes
stiffer, then the mixing angle is smaller and does not vary so rapidly
as a function of the C.M. coordinate, i.e. there is no
Landau-Zener crossing any more. For this reason one expects the
adiabatic limit to be recovered for a stiff object. To illustrate that we
 show in Fig.~7 the transmission coefficient obtained from the coupled
channel calculation, the no coupling case and the adiabatic one for $\hbar\Omega=3$ MeV.
\begin{figure}
\resizebox*{8cm}{5.5cm}{\includegraphics[angle=0]{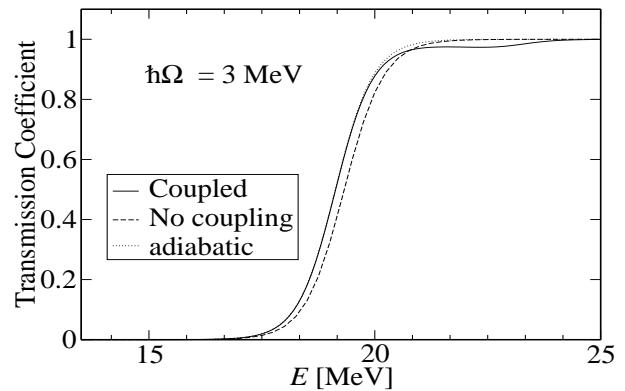}}
\caption{Comparison
  of the coupled channel result with the no coupling
  calculation and with the adiabatic transition for $\hbar\Omega=3$ MeV.}
\end{figure}
 One can note that for energies lower than the external barrier the adiabatic
limit is now recovered (dotted curve lies on top of the solid
line). This time,  $\frac{\hbar^2}{2M}(\theta'_{max})^2 \approx 0.018$
MeV and  $\frac{\hbar^2}{2M}\theta''_{max} \approx 0.19$ MeV and thus the
changes in the kinetic energy matrix (\ref{kintras}) are negligible.
 The coupling to the excited state still leads to a higher
 penetration factor with respect to the uncoupled case.
For even stiffer systems,
 for example $\hbar\Omega=5$ MeV, we observe that  the  coupled
calculation  finally coincides also with  the uncoupled  and the
adiabatic one.
We do not obtain an energy shift of the coupled result with respect to
the uncoupled case as in \cite{paper2}, since in our consistent
treatment  the excitation energy and the potential matrix element are
not independent from each other: with growing $\Omega$ the excitation
energy increases but the coupling potential $V_{02}(R)$ goes to zero.
When the object is too stiff to
be excited  no structure is
found in the transmission coefficient: the internal system starts from
the ground state and emerges still in the ground state at the end of
the barrier. In this limiting case  a coupled channel
calculation is not necessary, since the internal degrees
of freedom do not play any role.

To summarize, 
we present a  schematic model to describe the
tunnel effect of a two-body system in a two level 
approximation. 
%The very same derivation  can be used for example to
%describe s-wave fusion cross section in a  three-dimensional world.
Assuming an harmonic oscillator as
internal interaction and a Gaussian external barrier one can give a
consistent description of the potential felt by the macroscopic
coordinate due to presence of internal degrees of freedom.
No  a priori parameterization of the  potential matrix elements  is assumed.
Therefore, in the coupled channel picture
the dynamics of the internal degrees of freedom and their interaction
with the external barrier are treated consistently. A stiffer system
not only has a larger intrinsic excitation energy but also the
potential matrix elements change accordingly.
The coupled channel calculation
shows a peaked structure in the transmission coefficient, that
accounts for the excitation of the intrinsic system.
We show that  for a soft object the mixing
angle is large and
changes rapidly sign, so that the adiabatic limit is not approached at
low energies.
The resonant transition to the excited  state is 
 explained  by  a Landau-Zener level-crossing in a non adiabatic picture. As expected, the adiabatic limit
is recovered  in case of a stiff object, where the energy is not
sufficient to excite the internal structure. 
The model results suggest to investigate more carefully fusion of soft
nuclei or electron screening at astrophysical
energies, where often the adiabatic approximation is used.

\vfill\eject

\end{document}